\documentclass[aps,pra,twocolumn,superscriptaddress,floatfix,longbibliography]{revtex4}
\usepackage{amsfonts}
\usepackage{amssymb}
\usepackage{graphicx}
\usepackage{dcolumn}
\usepackage{bm}
\usepackage{amsmath}
\usepackage[colorlinks,linkcolor=magenta,citecolor=blue,urlcolor=blue]{hyperref}
\usepackage{changes}
\begin{document}
	
\preprint{This line only printed with preprint option}

\title{Emergent entanglement phase transitions in non-Hermitian Aubry-Andr\'e-Harper chains}

\author{Shan-Zhong Li}
\affiliation{Department of Physics, Fuzhou University, Fuzhou 350116, Fujian, China}
\affiliation {Key Laboratory of Atomic and Subatomic Structure and Quantum Control (Ministry of Education), Guangdong Basic Research Center of Excellence for 
Structure and Fundamental Interactions of Matter, School of Physics, South China Normal University, Guangzhou 510006, China}

\affiliation {Guangdong Provincial Key Laboratory of Quantum Engineering and Quantum Materials, Guangdong-Hong Kong Joint Laboratory of Quantum Matter, Frontier Research Institute for Physics, South China Normal University, Guangzhou 510006, China}

\author{Xue-Jia Yu}
\email{xuejiayu@fzu.edu.cn}
\affiliation{Department of Physics, Fuzhou University, Fuzhou 350116, Fujian, China}
\affiliation{Fujian Key Laboratory of Quantum Information and Quantum Optics,
College of Physics and Information Engineering,
Fuzhou University, Fuzhou, Fujian 350108, China}

\author{Zhi Li}
\email{lizphys@m.scnu.edu.cn}
\affiliation {Key Laboratory of Atomic and Subatomic Structure and Quantum Control (Ministry of Education), Guangdong Basic Research Center of Excellence for 
Structure and Fundamental Interactions of Matter, School of Physics, South China Normal University, Guangzhou 510006, China}

\affiliation {Guangdong Provincial Key Laboratory of Quantum Engineering and Quantum Materials, Guangdong-Hong Kong Joint Laboratory of Quantum Matter, Frontier Research Institute for Physics, South China Normal University, Guangzhou 510006, China}
\date{\today}

\begin{abstract}
We investigate the entanglement dynamics of the non-Hermitian Aubry-Andr\'e-Harper (AAH) chain. The results reveal that by increasing quasiperiodic strength, a phase transition occurs from the area law induced by non-Hermitian skin effect to the area law arising from Anderson localization. For the former, the entanglement entropy follows a non-monotonic process, i.e., it increases first, then oscillates, and finally converges to a stable value. While for the latter, the entanglement entropy remains low because the wave function is not expandable in Anderson's localization region. The early-stage behavior of entanglement entropy indicates that the two area-law cases are of different phases. Interestingly, the volume-law behavior emerges at the critical point between these two area-law phases. Our study reveals that the area laws induced by the skin effect and the Anderson localization is two different phases, and that a volume law can emerge at the phase transition point. The understanding of the entanglement phase transition induced by disorder and skin effect is thus deepened.
\end{abstract}

\maketitle

\section{Introduction}
Rapid development of experimental platforms for quantum simulations drives exploration of non-equilibrium dynamics~\cite{IMGeorgescu2014, RBlatt2012, CSchneider2012, JEisert2015, TLangen2015, CGross2017}. In recent years, significant attention has been directed towards the non-Hermitian Hamiltonian~\cite{YXiong2018,ZGong2018,SYao2018,KKawabata2019,CHLee2019,KYokomizo2019,LXiao2020, KZhang2020, NOkuma2020, DSBorgnia2020, ZYang2020, CXGuo2021, LLi2020, EJBergholtz2021,KKawabata2023,HJiang2019,VVKonotop2016,RElGanniny2018,YAshida2020,CMBender1998,AGuo2009,ARegensburger2012,SWeimann2017,XNi2018,MKremer2019,SXia2021,YLi2022,PPeng2016,JLi2019,ZRen2022,LXiao2017,HZLi2023,XJYu2023,ZXGuo2022,LZhou2023a,KLi2023,RHamazaki2019,Torito2023,YLGal2023,XTurkeshi2023,XZhang2022,DWZhang2020a,DWZhang2020b,LJLang2021,ZGong2017}, as they effectively describe the physical characteristics of open and non-conservative systems. One of the unique phenomena is the famous ``non-Hermitian skin effect'', and this asymmetric hoppings-induced effect can make the spectrum and eigenstates very much sensitive to boundary conditions.~\cite{SYao2018}. The skin effect holds a pivotal position in the realm of non-Hermitian topological band theories~\cite{YXiong2018,ZGong2018,SYao2018,KKawabata2019,CHLee2019,KYokomizo2019,LXiao2020, KZhang2020, NOkuma2020, DSBorgnia2020, ZYang2020, CXGuo2021, LLi2020, EJBergholtz2021}. Under open boundary condition (OBC), eigenstates become exponentially localized at the system's boundaries, disrupting the conventional bulk-boundary correspondence. Consequently, non-Hermitian bulk-boundary correspondence~\cite{ZGong2018,KZhang2020} and the non-Bloch band theory~\cite{SYao2018,KYokomizo2019} are invoked to account for the unconventional behavior induced by the skin effect. In the context of OBCs, the skin effect confines the macroscopic particle flow to the boundaries~\cite{KKawabata2023}, restraining the growth of entanglement entropy and adhering to the ``area law~\cite{JEisert2010}''.

In addition, Anderson localization also contributes to the suppression of entanglement entropy growth~\cite{RNandkishore2015,RNilanjan2021,Torito2023,KLi2023}. When disorder or quasiperiodic exceeds a threshold, the system undergoes a transition from an extended phase to a localized phase~\cite{PWAnderson1958,Fevers2008,ALagendijk2009,EAbrahams2010,SAubry1980}. Although both skin effect and Anderson localization can induce localized states, their eigenstate characteristics are completely different, i.e., the former is localized on the boundary, while the latter is localized in the bulk of the system. In low-dimensional ($d\le2$) systems, even minor disorder can drive the system into a localized phase, rendering the extended-localized transition absent in such cases~\cite{EAbrahams1979}. In three-dimensional case, weak disorder can lead to the coexistence of extended and localized states, which is clearly divided by the critical energy mobility edge~\cite{Fevers2008,SZLi2023a}. In other words, under such circumstance, one can achieve the metal-insulator phase transition by manipulating the Fermi surface.

Replacing random disorder with a quasiperiodic potential has been demonstrated to induce a metal-insulator transition~\cite{SAubry1980} and the mobility edge~\cite{XDeng2019,NRoy2021,JBiddle2010,SGanshan2015,YWang2020,SZLi2023,APadhan2022,SLellouch2014,HYao2019,TLiu2022} in low-dimensional systems. The one-dimensional AAH model is one of the most notable quasiperiodic examples~\cite{SAubry1980,PGHarper1955}. When one introduce asymmetric hopping to the AAH model, the extended phase gradually transforms into a skin localized phase under OBCs. The increase of quasi-periodic intensity weakens the skin effect, and the phase transition from skin state to Anderson localized state occurs gradually~\cite{HJiang2019}. The related dynamics have been extensively studied~\cite{LJZhai2020,SLonghi2021,LJZhai2022,TOrito2022,Torito2023,AChakrabarty2023,YCWang2023}. In addition, the change of entanglement entropy is accompanied by the phase transition from the skin phase to the localized phase. Then how does the entanglement behavior change before and after the phase transition? Do they vary depending on the boundary conditions one choose? What are the characteristics of the critical points between them?

To address the above questions, we investigate the entanglement dynamics of the AAH model with asymmetric (non-reciprocal) hopping. Under OBCs, Anderson localization is accompanied by a re-entrant area law ($S\propto L^{d-1}$) in entanglement entropy, i.e., transitioning from an area law induced by the skin effect to another area law induced by Anderson localization. The two area laws induced by these two different mechanisms can be distinguished by the early stages entanglement dynamics. Entanglement entropy under the skin effect exhibits initial growth, followed by a period of oscillation, and then gradual reduction. The non-monotonic growth observed in the early stages of evolution, as revealed through the density distribution, is driven by the skin effect, which forces particles to move along a unidirection. We demonstrate through early stage of entanglement dynamics and density evolution with different initial states that the time it takes for particles to accumulate at the boundary corresponds to the time it takes for entanglement entropy to decrease to a stable value. In contrast, under Anderson localization, the particles are localized at the initial position, which makes the entanglement entropy hardly grow. This early behavior is clearly illustrated by two different area law phases. Interestingly, by observing the scaling behavior at critical line, we find that a volume law ($S\propto L^{d}$) can emerge between the two different types of area law regions. Furthermore, we also discuss the entanglement phase transition under PBCs. As the quasiperiodic strength increases, a log-area law entanglement phase transition occurs, which again satisfies the volume law at its critical point.

The organization of this paper is as follows. In Sec.~\ref{II}, we introduce the model and outline the method for calculating the entanglement entropy. In Sec.~\ref{III}, we use the evolution of the early entanglement entropy and density distributions to distinguish between two different area law regions. In Sec.~\ref{IV}, we investigate the scaling behavior of the critical point between two area law regions in the non-Hermitian AAH model. In Sec.~\ref{V}, we discuss the case of PBCs. In Sec.~\ref{VI}, we presents a summary of the entire work. In Appendix~\ref{A}, we give more details on computing entanglement entropy. In Appendix~\ref{C}, We discuss the localization phase transition of the system and the multifractal properties on the critical line. In Appendix~\ref{B}, We give the variation of entanglement entropy with system size for small asymmetric hopping strengths.

\section{Model and entanglement entropy}\label{II}
We consider a non-Hermitian AAH model with asymmetric hopping and the Hamiltonian can be defined as
\begin{equation}\label{Hami}
H=\sum_{j=1}^{L-1}(J_{L}c_{j}^{\dagger}c_{j+1}+J_{R}c_{j+1}^{\dagger}c_{j})+\sum_{j=1}^{L}2\lambda\cos(2\pi\alpha j+\theta)c_{j}^{\dagger}c_{j},
\end{equation}
where $c_{j}$ ($c_{j}^{\dagger}$) is the fermionic annihilation (creation) operator at the $j$-th site, $L$ is the total number of the lattice. $J_{L}=-(J-\gamma)/2$ and $J_{R}=-(J+\gamma)/2$ with $J$ and $\gamma$ being real parameters depicting the strengths of symmetric and asymmetric hopping, respectively. The on-site potential strength is governed by the quasiperiodic strength $\lambda$. $\alpha$ is the quasiperiodic parameter, and $\theta$ is a random phase. When $\gamma=0$, $\lambda\neq 0$, Eq.~\eqref{Hami} corresponds to the AAH model, with a localization transition point at $\lambda=0.5$~\cite{SAubry1980}. When $\lambda=0$, $\gamma\neq 0$, Eq.~\eqref{Hami} reduces to the Hatano-Neslon model~\cite{NHatano1996}, which has skin effects under OBCs and has been extensively studied in the field of non-Hermitian topological insulator~\cite{YXiong2018,ZGong2018,SYao2018,KKawabata2019,CHLee2019,KYokomizo2019,LXiao2020, KZhang2020, NOkuma2020, DSBorgnia2020, ZYang2020, CXGuo2021, LLi2020, EJBergholtz2021}. This asymmetric hopping can be realized by the quantum trajectory approach~\cite{ZGong2018,KKawabata2023}. When $\gamma\neq 0$ and $\lambda\neq 0$, the localization transition point is given by $\lambda=\max\left\{|J_{R}|,|J_{L}|\right\}$~\cite{HJiang2019}. All eigenstates at the OBC (PBC) before reaching the critical quasiperiodic strength are skin states (extended states), and beyond the critical strength, they become Anderson localized states. Details about Anderson localization can be found in Appendix~\ref{C}. In the following, we will set $J = 1$ as the units of energy, $\alpha=(\sqrt{5}-1)/2$, $\lambda>0$ and $\gamma<0$.

To study the dynamical behavior of entanglement entropy, we choose the Ne\'el state $\left|\psi_{0}\right>=  {\textstyle \prod_{j=1}^{L/2}}c_{2j}^{\dagger}\left| \mathrm{vac} \right> $ as the initial state, where $\left| \mathrm{vac} \right> $ is the fermionic vacuum state. The initial wave function evolves according to the non-Hermitian Hamiltonian $H$ as
\begin{equation}\label{evl}
\left|\psi(t)\right>=\frac{e^{-iHt}\left|\psi_{0}\right>}{\left \|e^{-iHt} \left|\psi_{0}\right> \right \| }.
\end{equation}
Since the Hamiltonian in Eq.~\eqref{Hami} is quadratic and the initial state $\left|\psi_{0}\right>$ is a Slater determinant state, the final state $\left|\psi(t)\right>$ is also a determinant state and its correlation matrix $C_{ij}(t)=\left<\psi(t)\right|c_{i}^{\dagger}c_{j} \left|\psi(t)\right>$ can be efficiently calculated. The von Neumann entanglement entropy $S$ between a subsystem $\left[x_{1},~x_{2}\right]$ and the rest of the system
by~\cite{{IPeschel2003}}
\begin{equation}\label{evl}
S=-\sum_{i=1}^{x_{2}-x_{1}+1}\left[V_{i}\log(V_{i})+(1-V_{i})\log(1-V_{i})\right],
\end{equation}
where $V_{i}$ is the $i$th eigenvalue of the correlation matrix $C$. In the next calculations, we consider only the half-chain entanglement entropy, i.e., $x_1=1$ and $x_2=L/2$, and denote this as $S_{L/2}$ (More details can be found in Appendix~\ref{A}). We note that in the numerical calculations, all quantities are averaged over 300 random quasiperiodic phases $\theta$.

\section{Early state Entanglement dynamics}\label{III}

\begin{figure}[tbp]
\centering 
\includegraphics[width=8.5cm]{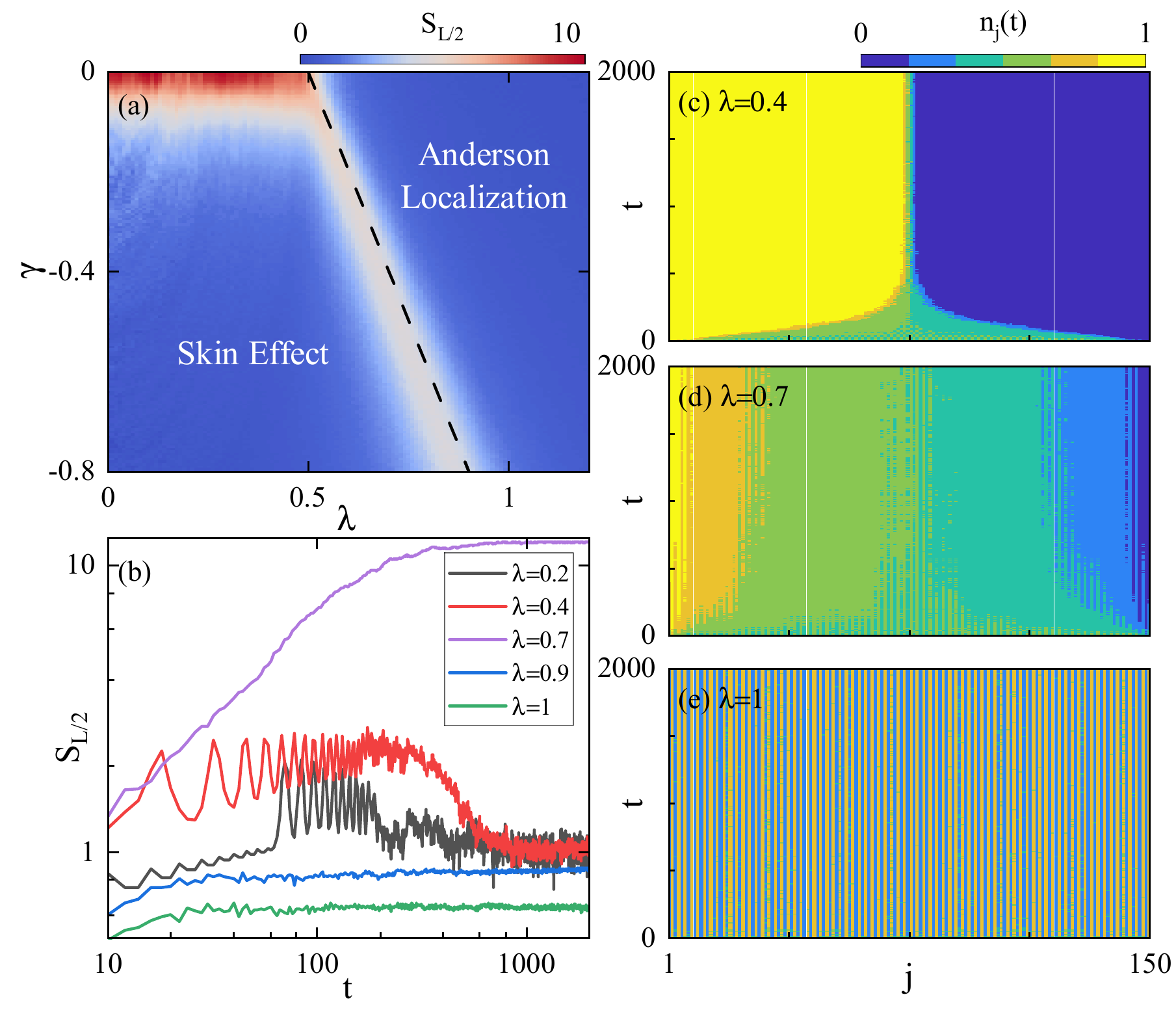} 
\caption{(a) Steady-state half-chain entanglement entropy $S_{L/2}$ on the $\lambda-\gamma$ plane with the system size $L=50$, where the black dashed line is $\lambda=(J-\gamma)/2$. (b) Time evolution of entanglement entropy $S_{L/2}$ for different $\lambda$. Time evolution of the density distribution $n_{j}(t)$ for (c) $\lambda=0.4$, (d) $\lambda=0.7$, and (e) $\lambda=1$. For the plot (b)-(e), $\gamma=-0.4$ and $L=150$. }
    \label{F1}
\end{figure}

In this section, we discuss the entanglement phase transitions for non-Hermitian AAH models under OBCs. The half-chain entanglement entropy $S_{L/2}$ of the steady-state in the $\lambda-\gamma$ plane is shown in Fig.~\ref{F1}(a), where the black dashed line is the Anderson localization transition line $\lambda=(J-\gamma)/2$~\cite{HJiang2019}. In the case of $\gamma=0$, as $\lambda$ crosses the critical value of 0.5, the system undergoes an entanglement transition from a volume law to an area law due to the influence of Anderson localization, with the critical point exhibiting a volume law~\cite{RNilanjan2021}. Clearly, the growth of entanglement entropy on both sides of the critical line is suppressed for $\gamma < 0$, but this is attributed to different mechanisms. On the left side of the critical line, when $\gamma \neq 0$, the skin effect propels particles towards the boundary, ultimately leading to particle localization at the boundary, reducing quantum jumps, and thus inhibiting the growth of entanglement entropy~\cite{KKawabata2023}. At the same time, we note that when $\gamma\rightarrow 0$, the system can obtain larger entanglement entropy (compared to the region with $\gamma < -0.3$), and we show that it remains an area law with further discussion in Appendix~\ref{B}. On the right side, Anderson localization results in exponential particle localization at the initial positions, suppressing the growth of entanglement entropy~\cite{RNandkishore2015,RNilanjan2021,Torito2023,KLi2023}. Both mechanisms suppress entanglement entropy, causing the system to follow the area law~\cite{RNandkishore2015,KLi2023,RNilanjan2021,Torito2023,KKawabata2023}.

In fact, the area law regions induced by these two different mechanisms can be distinguished by their early stage entanglement dynamics. In Fig.~\ref{F1}(b) we show the evolution of the early stage entanglement entropy $S_{L/2}$ for different $\lambda$ with $\gamma = -0.4$. For $\lambda<0.7$, the entanglement entropy initially increases, undergoes oscillations for a period, and then gradually decreases. This non-monotonic time evolution of entanglement entropy has been reported in many studies~\cite{RHamazaki2019,LJZhai2020,APanda2020,TOrito2022,Torito2023,KKawabata2023}. For $\lambda>0.7$, it consistently maintains a lower entanglement entropy. At the critical point $\lambda=0.7$, the entanglement entropy monotonically increases and then reaches a large saturation value. The distinction between these two area laws can also be characterized through the evolution of the density distribution $n_{j}$, defined as
\begin{equation}
n_{j}(t)=\left<\psi(t)\right|c_{j}^{\dagger}c_{j}\left|\psi(t)\right>.
\end{equation}
The evolution of the particle density distribution for $\lambda = 0.4$, $\lambda=0.7$, and $\lambda=1$ with the system size $L = 150$ is shown in Fig.~\ref{F1}(c)-(e). For $\lambda=0.4$, the skin effect pushes the particles in the direction imposed by the asymmetry, leading to an increase in early stage entanglement entropy. This phenomenon continues until the particles gradually localize at the boundary, before decreasing to a stable value. When $\lambda=1$, Anderson localization restricts particle transport, resulting in no increase in the entanglement entropy during both short and long-time evolution, always maintaining the information from the initial moment. For the critical point $\lambda=0.7$, the long-time propagation of particles causes the entanglement entropy to grow.

\begin{figure}[tbp]
\centering 
\includegraphics[width=8.5cm]{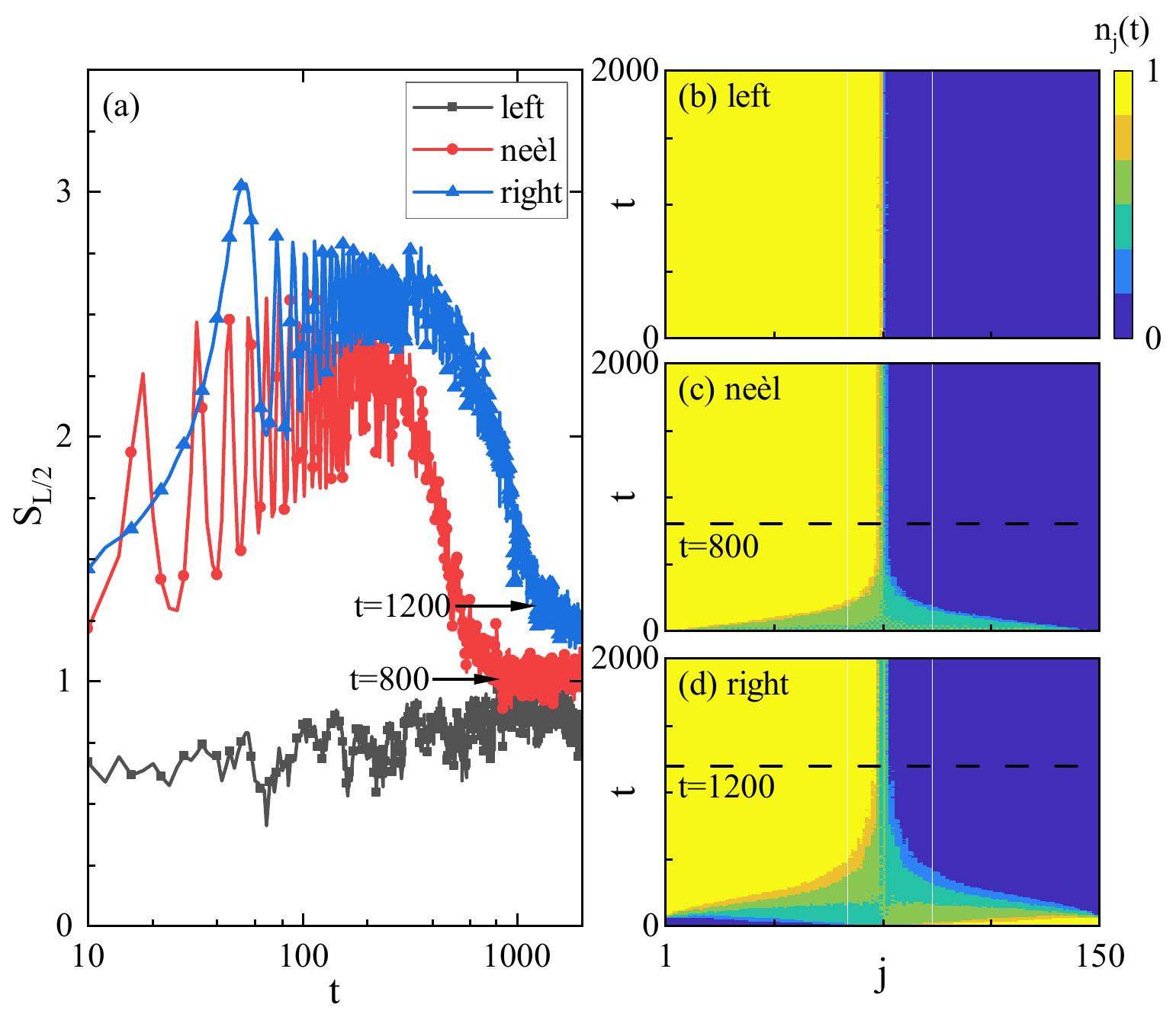} 
\caption{(a) Time evolution of entanglement entropy $S_{L/2}$ for different initial states. Time evolution of the density distribution $n_{j}(t)$ in the initial state (b) left, (c) ne\'el, and (d) right. For all plots, $\gamma=-0.4$, $\lambda=0.4$, and $L=150$.}
    \label{F2}
\end{figure}

In addition, the density evolution and the entanglement entropy reach a stable value at the same time. We consider other two initial states, localized in the left and right half chains, defined respectively as
\begin{equation}
\begin{aligned}
&\left|\mathrm{left}\right>= {\textstyle \prod_{j=1}^{L/2}}c_{j}^{\dagger}\left| \mathrm{vac} \right>,\\
&\left|\mathrm{right}\right>= {\textstyle \prod_{j=1}^{L/2}}c_{L/2+j}^{\dagger}\left| \mathrm{vac} \right>,
\end{aligned}
\end{equation}
and $U_{0}$ corresponding to these two states are $\left [U_{0}  \right ]_{jk}=\delta_{j,k}$ and $\left [U_{0} \right ]_{jk}=\delta_{L/2+j,k}$, respectively [for details see Appendix~\ref{A}]. Together with the ne\'el state the time evolution of the density distribution $n_{j}(t)$ for the three different initial states at $\lambda = 0.4$ is shown in Fig.~\ref{F2}(a). For the left state (localized in the left half chain), the entanglement entropy behaves similarly to that of the Anderson localized, maintaining relatively low values throughout. However, for the Ne\'el state and the right state (localized in the right half chain), the entanglement entropy exhibits an initial increase, followed by oscillations, and eventually decreases. We have provided approximate times required for the initial Ne\'el state and right state to reach a steady state in the figures, which are $t=800$ and $t=1200$, respectively. By comparing this with the density distribution in Fig.~\ref{F2}(b)-(d), one can see that the initial position of the left state lies on the skin boundary, so that the particle does not diffuse and therefore the entanglement entropy does not increase, exhibiting a behavior similar to that of Anderson localization. Conversely, for the initial Ne\'el and right states, the skin effect drives particle propagation towards the skin boundary, and the localization is at the boundary for the same time as the entanglement entropy reaches a stable value.

\section{Entanglement phase transition}\label{IV}
\begin{figure}[tbp]
\centering 
\includegraphics[width=8.5cm]{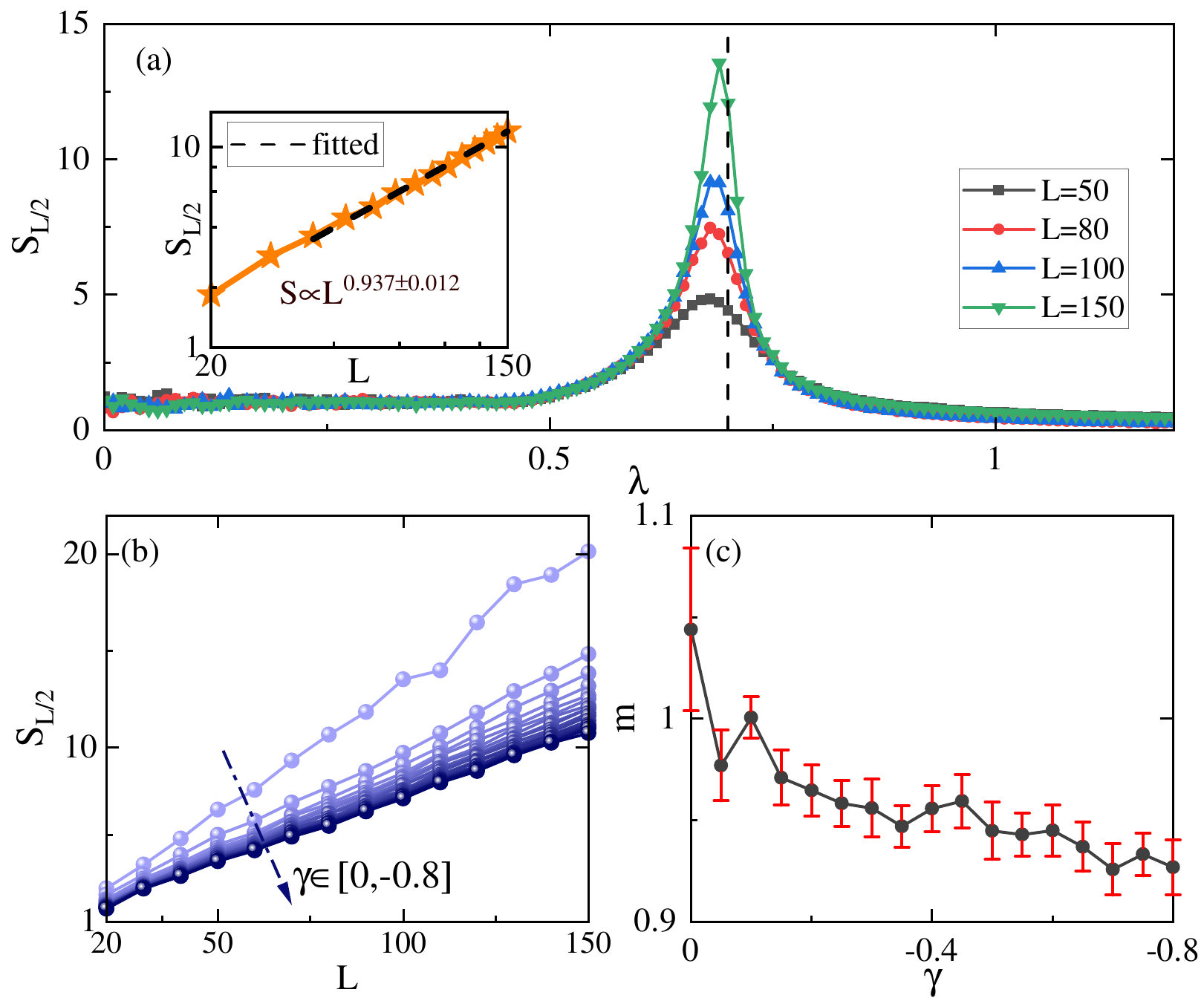} 
\caption{(a) Entanglement entropy $S_{L/2}$ as a function of $\lambda$ for different $L$ with $\gamma=-0.4$. The inset shows the data for $\lambda=0.7$ in a log-log scale with a power-law fit described by $S_{L/2}\propto L^{0.937\pm 0.012}$ and the black dashed line represents the transition point between the two area law regions with $\lambda=0.7$. (b) Entanglement entropy as a function of $L$ for different $\gamma$, where the darker the color the smaller the $\gamma$, and $\lambda=(J-\gamma)/2$. (c) The exponent $m$ fitted as a function of $\gamma$ in the form $S\propto L^{m}$.}
    \label{F3}
\end{figure}

The early entanglement dynamics provide a clear distinction between these two area law regions, indicating different phases. Naturally, the interesting question arises, is there an entangled phase transition between two different area law regions? Do the phase transition points still follow the area law? The time evolution of entanglement entropy at the critical point $\lambda=0.7$ can reach a relatively large saturation value, which suggests that a new law emerges at the critical point. In Fig.~\ref{F3}(a), we present the entanglement entropy $S_{L/2}$ as a function of $\lambda$ for various system sizes $L$ with $\gamma=-0.4$. It can be observed that at the critical point $\lambda=J_{L}=0.7$, the entanglement entropy increases with the enlargement of the system size, while the two regions governed by the area laws remain constant. In the inset, we fit the scaling behavior of the entanglement entropy at the critical point, which satisfies the nearly volume law $S_{L/2}\propto L^{0.937\pm 0.012}$. That is, similar to the Hermitian case, the volume law on the critical point is robust even in the presence of skin effect, which stems from the fact that the critical point remains a multifractal phase under the influence of asymmetry hopping, as discussed in Appendix~\ref{C}. Further, to verify whether the volume law is maintained at the critical line, we show the entanglement entropy as a function of $L$ for $\gamma$ from 0 to -0.8 in Fig.~\ref{F3}(b), where the darker the color the smaller the $\gamma$, and $\lambda=(J-\gamma)/2$. It is clear that the introduction of asymmetry hopping suppresses the steady state entanglement entropy. However, in Fig.~\ref{F3}(c), where we used a power law form $S_{L/2}\propto L^m$ to fit the scaling of entanglement entropy, one can observe that $m>0.9$ almost follows the volume law. This implies that the scaling behavior at the critical line is robust with respect to asymmetric hopping, and the volume law can emerge between the two area law regions, leading to an area-area entanglement phase transition.

\section{periodic boundary conditions}\label{V}
\begin{figure}[htbp]
\centering 
\includegraphics[width=8.5cm]{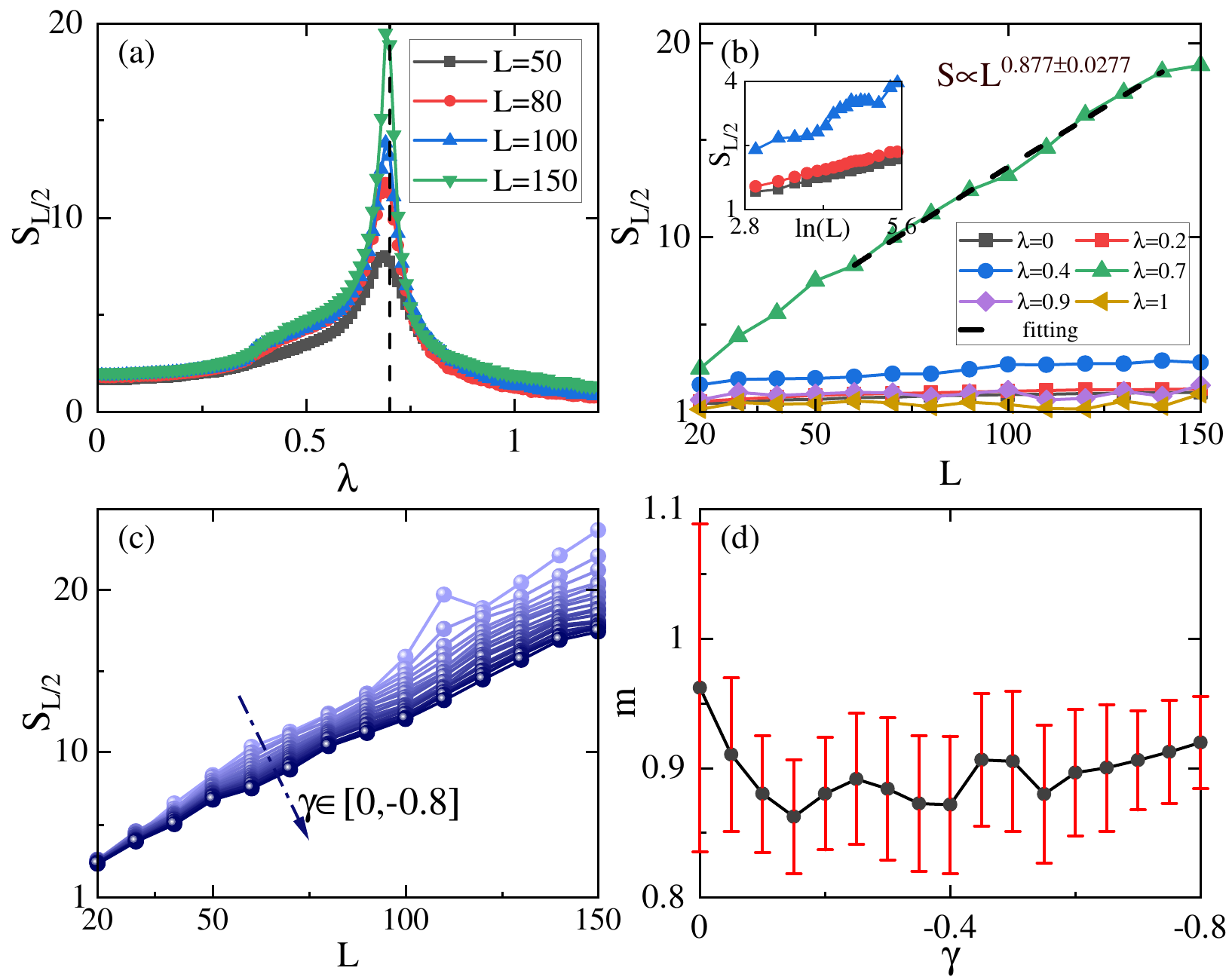} 
\caption{(a) Entanglement entropy $S_{L/2}$ as a function of $\lambda$ for different $L$ with $\gamma=-0.4$, where the black dashed line represents the transition point between the two area law regions with $\lambda=0.7$. (b) The linear-log plot of the entanglement entropy $S_{L/2}$ as a function of $L$ for different $\lambda$. The inset shows $S_{L/2}$ in log-linear coordinates for $\lambda= 0,0.2,0.4$ with $L$ up to 250. (c) Entanglement entropy as a function of $L$ for different $\gamma$, where the darker the color the smaller the $\gamma$, and $\lambda=(J-\gamma)/2$. (d) The fitting exponent $m$ as a function of $\gamma$.}
    \label{F4}
\end{figure}

In this section, we discuss the case of PBCs. In Fig.~\ref{F4}(a) we show the entanglement entropy $S_{L/2}$ as a function of $\lambda$ for different system sizes $L$. In the absence of quasiperiodic potential, it has been demonstrated in Ref.~\cite{KKawabata2023,KLi2023} that the entanglement entropy satisfies the logarithmic relation $S_{L/2}=\frac{1}{3}\log L$ with system size $L$. When $\lambda<0.7$, the entanglement entropy still satisfies the logarithmic law [see inset of Fig.~\ref{F4}(b)]. For $\lambda>0.7$, the Anderson localization makes the entanglement entropy follow the area law. However, at the critical point $\lambda=0.7$, the entanglement entropy increases as the system size increases, forming a distinct peak. In Fig.~\ref{F4}(b), we show the entanglement entropy as a function of $L$ for different $\lambda$. We fit $m$ on the critical point by $S\propto L^m$ and $m=0.877\pm 0.027$ still almost keeps the volume law. Furthermore, in Fig.~\ref{F4}(c), we present the entanglement entropy $S_{L/2}$ as a function of $L$ for critical line (the darker the color the smaller the $\gamma$), and the fitted values of $m$ are shown in Fig.~\ref{F4}(d). It can be observed that the introduction of asymmetric hopping $\gamma$ slightly suppresses the growth of entanglement but almost maintains the volume law. Under PBCs, the increase of the quasiperiodic strength induces an entanglement phase transition of the log-area law, which is the same conclusion as in Ref.~\cite{Torito2023}.

\section{conclusion}\label{VI}
In this paper, we focus on the entanglement dynamics of a non-Hermitian AAH chain with asymmetric hopping. We obtain the corresponding phase diagram for OBCs and PBCs, respectively. 

On the one hand, for OBCs, the entanglement entropy of the system is suppressed both before and after the phase transition, showing the characteristics of area law behavior. By further analyzing the entanglement dynamics and density distribution features in the early stage of evolution, we find that the system of the above two area law belongs to different phases, which can be distinguished by the entanglement entropy. The difference between the two area law phases means that a phase transition occurs between them. Note that, through finite-size scaling analysis, we find that the critical boundary of the phase transition exhibits the characteristics of volume law. 

On the other hand, for PBCs, the disappearance of skin effect will cause logarithmic law behavior before the localization phase transition, thus gives rise to log-area law entanglement phase transition in the system. Furthermore, a similar critical boundary with volume law characteristics can emerge. 

This work reveals that the area-law phases induced by disorder and non-Hermitian are essentially different phases, and the phase transition points between them show the characteristics of volume law. This opens a new avenue to explore entanglement phase transitions in non-Hermitian quasi-periodic disorder systems.

\emph{Note added:} In completing this manuscript, we note that a recent preprint with a similar title to ours is entitled "Entanglement phase transitions in non-Hermitian quasicrystals" [arXiv:2309.00924]~\cite{LZhou2023b}, which similarly investigated entanglement phase transitions in non-Hermitian AAH models. The paper revealed two entanglement phase transitions in non-Hermitian quasicrystals, the volume-area law, and the log-area law. Here, in our paper, area-area law entanglement phase transition under asymmetric hopping is discoverd, and a volume law can emerge between two different area law phase. Note that,the log-area law phase transition mentioned in the preprint is a special case of our model under the PBC.

\begin{acknowledgments}
We thank Shi-Liang Zhu, Yu-Guo Liu, and Enhong Cheng for helpful discussions and constructive suggestions. This work was supported by the Guangdong Basic and Applied Basic Research Foundation (Grants No.2021A1515012350) and the National Key Research and Development Program of China (Grant No.2022YFA1405300). X.-J.Yu acknowledges support from the start-up grant XRC-23102 of Fuzhou University.
\end{acknowledgments}

\appendix
\section{Details of the calculation of entanglement entropy}\label{A}
Here, we will give the details of computing the entanglement entropy (also see~\cite{KKawabata2023,KLi2023}). The dynamics of the initial ne\'el state $\left|\psi_{0}\right>=  {\textstyle \prod_{j=1}^{L/2}}c_{2j}^{\dagger}\left| \mathrm{vac} \right> $ at the free fermions Hamiltonian $H$ up to moment $t$ is given by
\begin{equation}
\left|\psi(t)\right>=\frac{e^{-iHt}\left|\psi_{0}\right>}{\left \|e^{-iHt} \left|\psi_{0}\right> \right \| },
\end{equation}
which can be further written as
\begin{equation}
\begin{aligned}
\left|\psi(t)\right>&=\frac{1}{\sqrt{N(t)}}e^{-iHt}\prod_{j=1}^{L/2}c_{2j}^{\dagger}\left|\mathrm{vac} \right>\\
&=\frac{1}{\sqrt{N(t)}}\prod_{j=1}^{L/2}c_{2j}^{\dagger}(t)\left|\mathrm{vac} \right>,
\end{aligned}
\end{equation}
where $N(t)=\left \|e^{-iHt} \left|\psi_{0}\right>   \right \| $, and $c_{j}^{\dagger}(t)=e^{-iHt}c_{j}^{\dagger}e^{iHt}$. We see that the evolved $\left|\psi(t)\right>$ remains a determinant, except that the operator $c_{2j}^{\dagger}(t)$ is not necessarily orthogonal. We can write the unnormalized evolving state as
\begin{equation}
\begin{aligned}
\left |\tilde{ \psi}  (t) \right \rangle &=\prod_{j=1}^{L/2}c_{2j}^{\dagger}\left |\mathrm{vac}  \right \rangle\\
&=\left [\prod_{k=1}^{L/2}\left (\sum_{j=1}^{L}\left [ U(t) \right ]_{jk}c_{j}^{\dagger}   \right )   \right ]  \left | \mathrm{vac}   \right \rangle ,
\end{aligned}
\end{equation}
where $U(t)=e^{-iHt}U_{0}$, and $U_{0}$ is an $L\times \frac{L}{2}$ matrix representing the set of all initial single-particle states with $\left [U_{0}  \right ]_{jk}=\delta_{j,2k}$. In this representation, the matrix $U(t)$ contains all information about the quantum dynamics. As the Hamiltonian is non-Hermitian, the elements in $U(t)$ may grow or decay exponentially with $t$. Therefore we need to determine smaller step sizes $\Delta t$ to prevent non-Hermitian instability. After the time interval $\Delta t$, the state evolves as
\begin{equation}
\begin{aligned}
\left|\psi(t+\Delta t)\right>&\propto e^{iH\Delta t}\left|\psi(t)\right>\\
&=\left [\prod_{k=1}^{L/2}\left (\sum_{j=1}^{L}\left [ e^{iH\Delta t}U \right ]_{jk}(t)c_{j}^{\dagger}   \right )   \right ]  \left | \mathrm{vac}   \right \rangle ,
\end{aligned}
\end{equation}
To restore the normalization condition $\left \langle \psi  | \psi  \right \rangle =1$, we perform the $QR$ decomposition,
\begin{equation}
     U(t)=e^{iHt}U_{0}=QR,
\end{equation}
where $Q$ is an $L\times \frac{L}{2}$ matrix satisfying $Q^{\dagger}Q=1$, and $R$ is an upper triangular matrix. The $L\times \frac{L}{2}$ matrix $U(t+\Delta t)$ is obtained as
\begin{equation}
U(t+\Delta t)=Q.
\end{equation}
For our calculations, $\Delta t \le  5$ was chosen to prevent non-Hermitian induced numerical instabilities.

The correlation function $C_{ij}(t)=\left<\psi(t)\right|c_{i}^{\dagger}c_{j}\left|\psi(t)\right> $ in time $t$, and the von Neumann entanglement entropy $S$ between a subsystem $\left[x_{1},~x_{2}\right]$ and the rest of the system
by~\cite{{IPeschel2003}}
\begin{equation}\label{evl}
S=-\sum_{i=1}^{x_{2}-x_{1}+1}\left[V_{i}\log(V_{i})+(1-V_{i})\log(1-V_{i})\right],
\end{equation}
where $V_{i}$ is the $i$th eigenvalue of the correlation matrix. In the main text we have only considered the half-chain, i.e. $x_1=1$, $x_2=L/2$. The density of site $j$ at time $t$ is
\begin{equation}\label{evl}
n_{j}=C_{jj}(t).
\end{equation}

\section{The localization phase transition}\label{C}
We can transform the Hamiltonian Eq.~\ref{Hami} with OBCs into the Hermitian AAH Hamiltonian $H'$ by similarity transformation
\begin{equation}
H'=SHS^{-1}=
\begin{pmatrix}
 V_{1} & J' &   & \\
 J' &  V_{2} & J' & \\
   & \ddots  & \ddots  &J' \\
  &  & J' &  V_{L}
\end{pmatrix},
\end{equation}
where $J'=\sqrt{J_{R}J_{L}}$, $V_{j}=2\lambda\cos(2\pi\alpha j +\theta)$, the similarity matrix $S=\mathrm{diag}(e^{-g},~e^{-2g},\dots,-e^{Lg})$ and $g=\sqrt{J_{R}/J_{L}}$. For the Hermitian's AAH model $H'$, $\lambda/J' = 1$ is the localization phase transition point. Let $\psi'$ be the eigenstate of the Hamiltonian $H'$, then for the eigenstate $\psi$ of the Hamiltonian $H$, it satisfies $\psi=S^{-1}\psi'$. Thus, for an extended eigenstate of the Hamiltonian $H'$, $S^{-1}$ makes the wave function exponential localized on the boundary. For a localized state, the wave function is 
\begin{equation}
    |\psi_{j}|\propto \left\{\begin{matrix}
e^{-(\eta+g)(j-j_{0})},& j>j_{0}, \\
e^{-(\eta-g)(j_{0}-j)},& j<j_{0},
\end{matrix}\right.
\end{equation}

where $j_{0}$ is the index of the localization center, and $\eta = \ln(\lambda/J') > 0$ is the Lyapunov exponent for Hamiltonian $H'$. There are two different Lyapunov exponents $\eta\pm g $ on both sides of the localized center. When $\eta\le |g|$, the system shows delocalized of Anderson mode and the emergent of skin mode on the same side, the corresponding boundary of the skin/localization phase is given by $\lambda=\max\left\{J_{L},J_{R}\right\}$~\cite{HJiang2019}.

Numerically, the transition from the skin phase to the Anderson localized phase can be described by the following physical quantities.

The first is the fractal dimension $\Gamma$. For the $\beta$th eigenstate $\left|\psi(\beta)\right>=\sum_{j}\psi_{j}(\beta)\left|j\right>$, and we may evaluate the moments $\xi_{q}(\beta)=\sum_{j=1}^{L}|\psi_{j}(\beta)|^{2q}\propto L^{-\Gamma_{q}(q-1)}$~\cite{HYao2019,XDeng2019,YWang2020}, where $\Gamma_{q}$ are the fractal dimensions. For the next calculation, we choose $q=2$ then the fractal dimension can be written as
\begin{equation}
\Gamma(\beta)=\lim_{L\rightarrow\infty}\frac{\ln\xi(\beta)}{\ln L}.
\end{equation}
We omit the subscripts for $\Gamma_{2}$ and $\xi_{2}$, then $\xi$ is the inverse participation ratio. For an extended (localized) state $\Gamma=1$ ($\Gamma= 0$), and $0<\Gamma<1$ for a critical state. Since there are no mobility edges in the non-Hermitian AAH model, we further define the average fractal dimension $\overline{\Gamma}=\frac{1}{L}\sum_{\beta=1}^{L}\Gamma(\beta)$ to account for the localization phase transition.

Second, the skin effect stems from the non-trivial topological properties of the system~\cite{HJiang2019,ZGong2018,NOkuma2020}. We can define the winding number of the reference point $E_{b}$ in the complex plane
\begin{equation}
\omega=\int_{0}^{2\pi} \frac{\partial_\Phi \ln \mathrm{det}[H(\Phi)-E_{b}]}{2\pi i}d\Phi,
\end{equation}
where $H(\Phi)=H+J_{R}e^{-i\Phi}+J_{L}e^{i\Phi}$ and $\Phi$ is a magnetic flux. When $\omega = 1~(-1)$, it has a left (right) skin phase under the OBC, while $\omega = 0$ has no skin effect. Moreover, under the skin phase, the Hamiltonian Eq.~\ref{Hami} is particularly sensitive to boundary conditions, and the complex eigenvalues vanish under OBCs. In the localized phase, the eigenvalues are all real independently of the boundary conditions~\cite{HJiang2019}. We can define the eigenvalue maximal imaginary part
\begin{equation}
\mathrm{max}(ImE)=\mathrm{max}_{\beta=1,\dots L}(|ImE_{\beta}|).
\end{equation}
to describe the skin-Anderson localization phase transition.

\begin{figure}[thbp]
\centering 
\includegraphics[width=8.5cm]{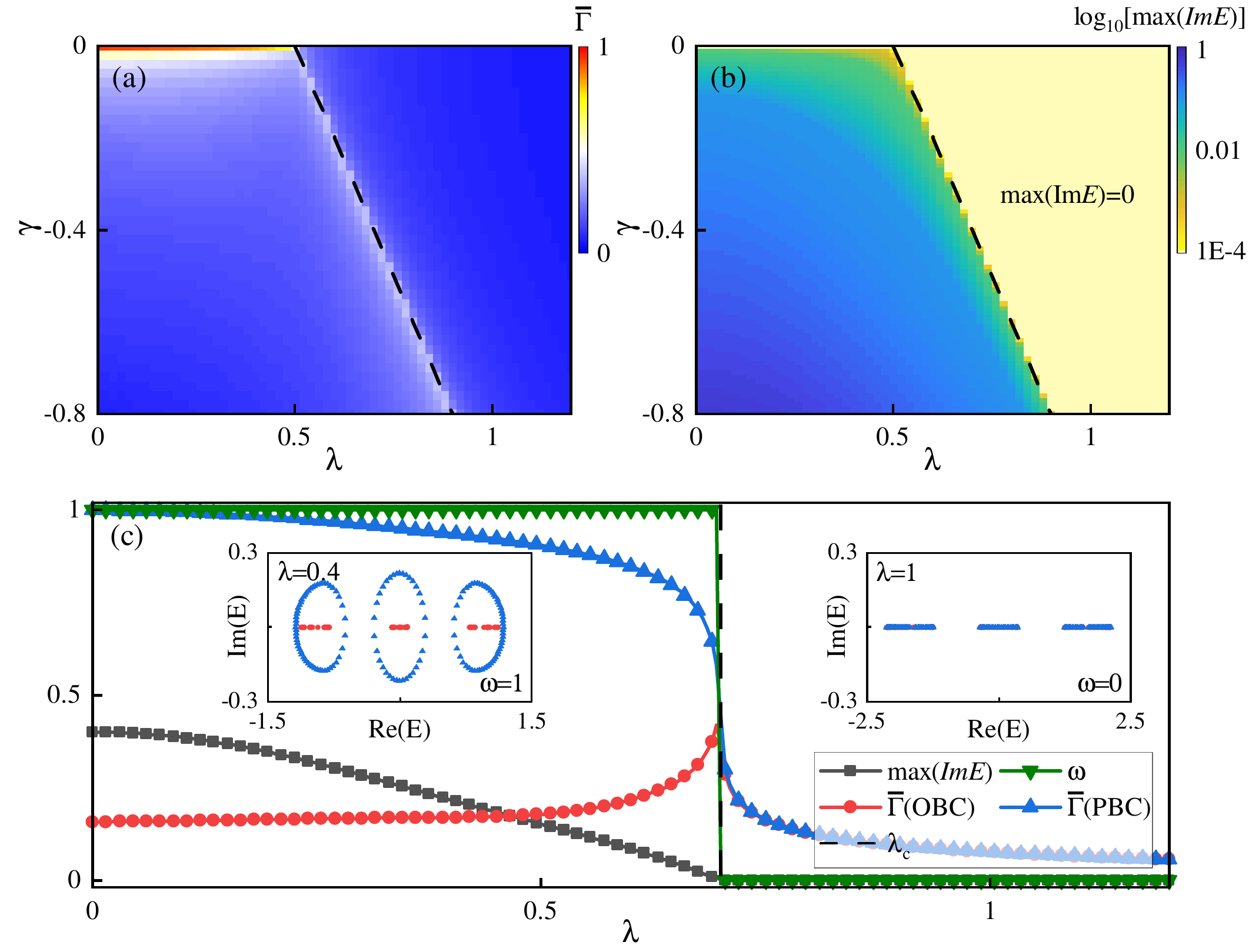} 
\caption{(a) The average fractal dimension $\overline{\Gamma}$ and (b) $\mathrm{max}(ImE)$ on the $\lambda-\gamma$ plane, where the black dashed line is $\lambda=(J-\gamma)/2$. (c) The average fractal dimension $\overline{\Gamma}$, $\mathrm{max}(ImE)$, and winding number $\omega$ as a function of $\lambda$ for $\gamma=-0.4$, where the inset shows the eigenvalues under PBC (blue triangles) and OBC (red circles) in the complex plane for system size $L=144$. For all main plots, we set $L=610$ and $\theta=0$.}
    \label{A6}
\end{figure}

In Fig.~\ref{A6}(a) and (b), we show the average fractal dimension $\overline{\Gamma}$ under OBC and $\mathrm{max}(ImE)$ under PBC, respectively. When $\gamma=0$, $\lambda=0.5$ is the phase transition point for the extended and localized regions. When $\gamma < 0$, $\mathrm{max}(ImE)>0$ shows that the system has a non-trivial topological point gap, in which case the eigenstates exhibit a skin effect under OBC, making $\Gamma$ tend to 0. However, at the critical line $\lambda=(J-\gamma)/2$, the $\overline{\Gamma}$ exhibits a peak, and the eigenvalue under PBC undergoes a complex-realistic transition, and the skin effect disappears. Specifically, in Fig.~\ref{A6}(c), we show $\mathrm{max}(ImE)$, $\overline{\Gamma}$, and $\omega$ as a function of $\lambda$ for $\gamma = -0.4$. The $\mathrm{max}(ImE)>0$ and $\omega=1$, showing that the system has a non-trivial topological point gap in the region $\lambda<\lambda_{c}$, causing $\overline{\Gamma}\rightarrow0$ under OBC due to the skin effect, while $\overline{\Gamma}\rightarrow 1$ under PBC is the extended phase. We can see the case of $\lambda=0.4$ in the inset in Fig.~\ref{A6}(c), where in the complex plane, the PBC eigenenergy spectrum contains the eigenvalues of the OBC, and the non-trivial winding number allows for a skin effect under OBC. When $\lambda>\lambda_c$, the non-trivial point gap vanishes, $\mathrm{max}(ImE),\omega=0$, and the system undergoes Anderson localization (the case of $\lambda=1$ in the inset). The system undergoes multiple phase transitions on the critical line $\lambda=\lambda_{c}$, i.e., localization phase transitions, eigenvalue complex-real phase transitions, and topological phase transitions.

\begin{figure}[tbp]
\centering 
\includegraphics[width=8.5cm]{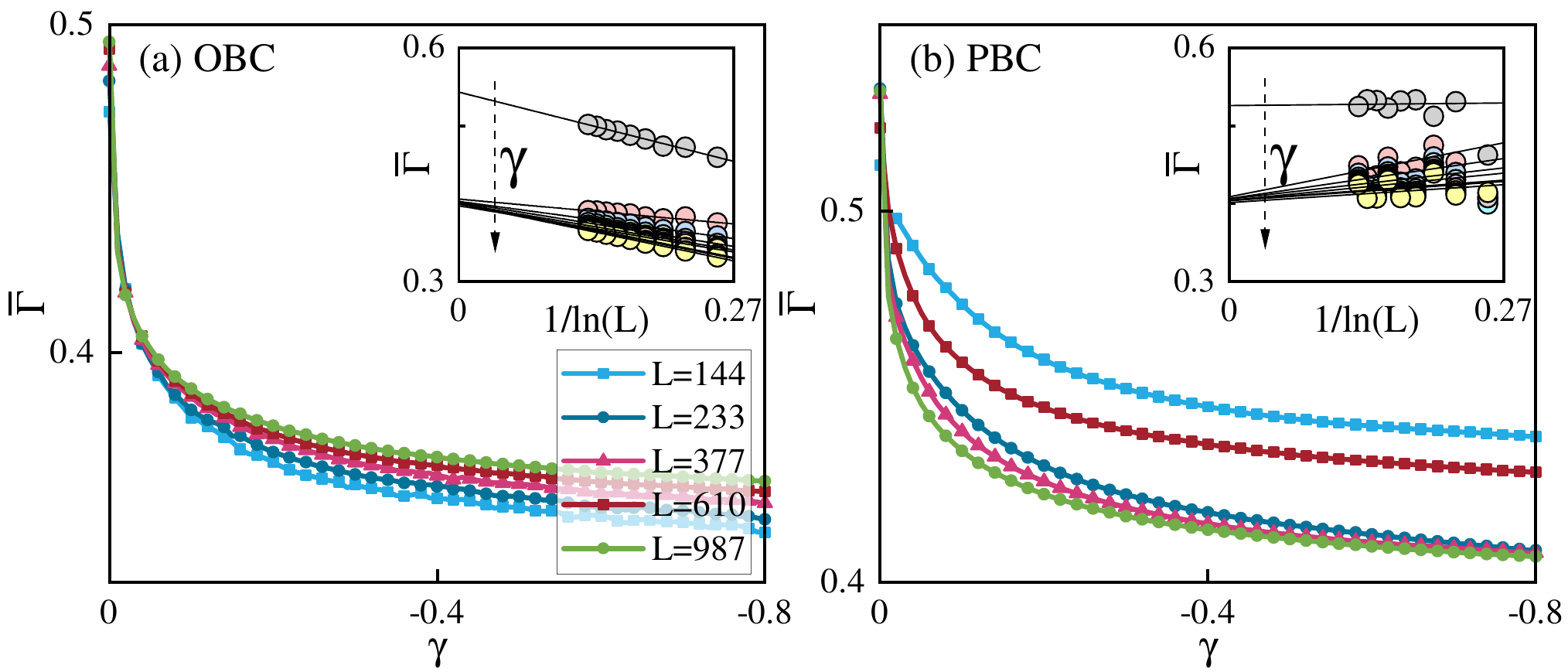} 
\caption{The average fractal dimension $\overline{\Gamma}$ as a function of $\gamma$ at the critical line under (a) OBC and (b) PBC, where $\lambda=(J-\gamma)/2$. The inset shows the $\overline{\Gamma}$ as a function of $\ln(L)$ for $\gamma$ from 0 to -0.8 and $\lambda=(J-\gamma)/2$. For all plots, we set $\theta=0$. }
    \label{A7}
\end{figure}
To understand the volume law that emerges at the critical line, we further discuss the effect of non-Hermitian hopping on the multifractal phase. In Fig.~\ref{A7}(a) and (b), we show the average fractal dimension $\overline{\Gamma}$ at the critical point under OBC and PBC, respectively. The results show that the increasing strength of non-Hermitian leads to decreasing $\overline{\Gamma}$ at the critical point, and this enhancement of localization is also reflected in a decrease in the steady-state entanglement entropy [see Figs.~\ref{F3}(b) and \ref{F4}(c)]. In addition, the $\overline{\Gamma}$ rises slightly with increasing system size under OBC, whereas it exhibits size-independence under PBC. We have calculated the $\overline{\Gamma}$ under $L = 55, ~89, ~144, ~233, ~377, ~610,~ 987, ~1597,~2584$ for $\gamma$ from 0 to -0.8 in the inset and interpolated to the thermodynamic limit $L\rightarrow\infty$ by linear fitting. It can be seen that in the thermodynamic limit, both for OBC and PBC, $\gamma<0$ at $\overline{\Gamma}\sim 0.4$, the system remains multifractal at the critical point, making the entanglement entropy behave as a volume law.

\section{Scaling behaviour of entanglement entropy under small $\gamma$}\label{B}
\begin{figure}[htbp]
\centering 
\includegraphics[width=8.5cm]{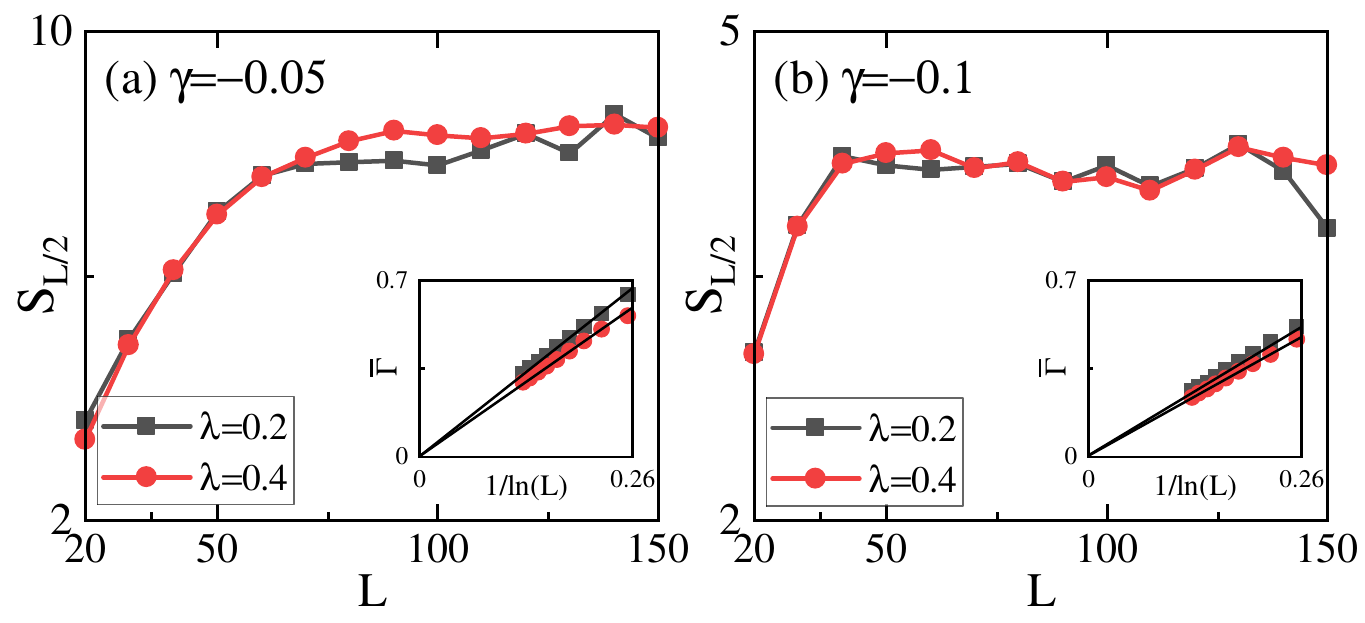} 
\caption{The entanglement entropy $S_{L/2}$ as a function of $L$ for different $\lambda$ with (a) $\gamma=-0.05$ and (b) $\gamma=-0.1$. The inset shows the $\overline{\Gamma}$ as a function of $\ln(L)$ for different $\lambda$.}
    \label{A5}
\end{figure}
In the main text, in Fig.~\ref{F1}(a), the system has a large value of entanglement entropy as $\gamma\rightarrow 0$ compared to $\gamma < -0.3$. This is due to the fact that the smaller strength of the asymmetric hopping makes the skin effect weaker and the entanglement entropy is able to grow to larger values. In Fig.~\ref{A5}(a) and (b), we respectively present the steady-state entanglement entropy as the system size increases for $\gamma = -0.05$ and $-0.1$. It can be observed that, at large system sizes, it satisfies the area law. Also in the inset we give the average fractal dimension $\overline{\Gamma}$ of the corresponding $\lambda$ at $L = 55, ~89, ~144, ~233, ~377, ~610,~ 987, ~1597,~2584$. When the system size is small, it has a large $\overline{\Gamma}$, and the system is less localized. As the size increases and is fitted to the thermodynamic limit, $\overline{\Gamma}=0$ indicates that it is a localized phase and the entanglement entropy still follows the area law.

\end{document}